\documentclass[]{spie}  

\usepackage{amsmath,amsfonts,amssymb}
\usepackage{subcaption}
\usepackage{graphicx}
\usepackage{cite} 
\usepackage{epsfig}
\usepackage{nccmath}
\usepackage{xcolor,colortbl}
\usepackage{tabularx}
\usepackage{relsize}
\usepackage{pifont}
\usepackage{booktabs} 
\usepackage{multirow}
\usepackage{multicol}
\usepackage{adjustbox}
\usepackage{float}
\usepackage{makecell}
\usepackage{tabu}
\usepackage[colorlinks=true, allcolors=blue]{hyperref}
\usepackage[capitalize]{cleveref}
\usepackage{algorithm}

\title{Dataset Distillation in Medical Imaging: A Feasibility Study}

\author[a]{Muyang Li}
\author[a]{Can Cui}
\author[a]{Quan Liu}
\author[a]{Ruining Deng}
\author[a]{Tianyuan Yao}
\author[a]{Marilyn Lionts}
\author[a]{Yuankai Huo}

\affil[a]{Vanderbilt University, Nashville TN 37235, USA}

\authorinfo{Further author information: (Send correspondence to Yuankai Huo)\\
E-mail: yuankai.huo@vanderbilt.edu}

\begin{document}
\maketitle

\begin{abstract}
Data sharing in the medical image analysis field has potential yet remains underappreciated. The aim is often to share datasets efficiently with other sites to train models effectively. One possible solution is to avoid transferring the entire dataset while still achieving similar model performance. Recent progress in data distillation within computer science offers promising prospects for sharing medical data efficiently without significantly compromising model effectiveness. However, it remains uncertain whether these methods would be applicable to medical imaging, since medical and natural images are distinct fields. Moreover, it is intriguing to consider what level of performance could be achieved with these methods. To answer these questions, we conduct investigations on a variety of leading data distillation methods, in different contexts of medical imaging. We evaluate the feasibility of these methods with extensive experiments in two aspects:  1) Assess the impact of data distillation across multiple datasets characterized by minor or great variations. 2) Explore the indicator to predict the distillation performance. Our extensive experiments across multiple medical datasets reveal that data distillation can significantly reduce dataset size while maintaining comparable model performance to that achieved with the full dataset, suggesting that a small, representative sample of images can serve as a reliable indicator of distillation success. This study demonstrates that data distillation is a viable method for efficient and secure medical data sharing, with the potential to facilitate enhanced collaborative research and clinical applications.
\end{abstract}


\section{Introduction}

Data plays a crucial role in machine learning and data analysis, with the importance of big data being demonstrated in many large-scale models today~\cite{l2017machine}. In medical environments, it is common to share image data between different hospitals for medical research, enhancing patient care, and facilitating the development of innovative treatments~\cite{hopfield1982neural}. Therefore, the topic of efficient data sharing between different sites is becoming increasingly important. Furthermore, given the sensitive nature of personal health information, ensuring the privacy and security of this data is crucial to maintaining patient trust and complying with legal and ethical standards~\cite{hulsen2020sharing, kaissis2020secure}.

Data distillation ~\cite{wang2020dataset} emerges as a practical solution, offering a means to share the crucial essence of the full dataset without the need to transfer the entire bulk. This approach aims to compress the original dataset into a much smaller dataset to increase the efficiency of model training and deployment, without sacrificing the model performance. Also, it avoids sharing the original data and is suitable for sharing-restricted scenarios. 

Among current dataset distillation methods, dataset condensation (DC)~\cite{zhao2021dataset} is a fundamental work, that firstly proposes to match the gradient between the original dataset and synthetic small dataset, which achieved good performance on natural image datasets. As a method developed from DC, MTT(Matching Training Trajectories) shows great performance on dataset distillation, speeding up the process and improving the general distillation work~\cite{chen2023comprehensive, cui2022dcbench}.

However, it remains uncertain whether the methods that work successfully on natural images would apply to medical imaging, which often features smaller class variability~\cite{2023arXiv230503711W}. Natural image data distillation methods were evaluated on ImageNet, CIFAR-10, CIFAR-100, and MNIST~\cite{beyer2020we, byerly2021routing}. These datasets, annotated by humans, feature images with relatively clear visual attributes that facilitate image classification. In contrast, medical images focus on preserving essential diagnostic details crucial for disease detection, often presenting subtler distinctions between classes~\cite{kumar2021integration, vu2024integrating}. Medical images typically involve similar biological tissues with subtle differences crucial for diagnosis, such as variations in tissue texture, density, or the presence of minute abnormalities. When examining small image patches from pathology or radiology images, distinguishing between benign and malignant tumors or different stages of a disease often requires expert knowledge to interpret subtle visual cues accurately. This necessitates a specialized approach that maintains critical medical information, distinct from the broader aims of natural image distillation~\cite{morra2021bridging}. Furthermore, the wider applicability of dataset distillation to various types of medical datasets has yet to be explored. A soft-label anonymous gastric X-ray image distillation method has been proposed~\cite{Li_2020, li2022compressed}, but their work was limited to a single modality. This significant difference in inter-class variation poses unique challenges for dataset distillation methods when applied to medical images.

\begin{figure}[!h]
    \centering
    \includegraphics[width=1\textwidth]{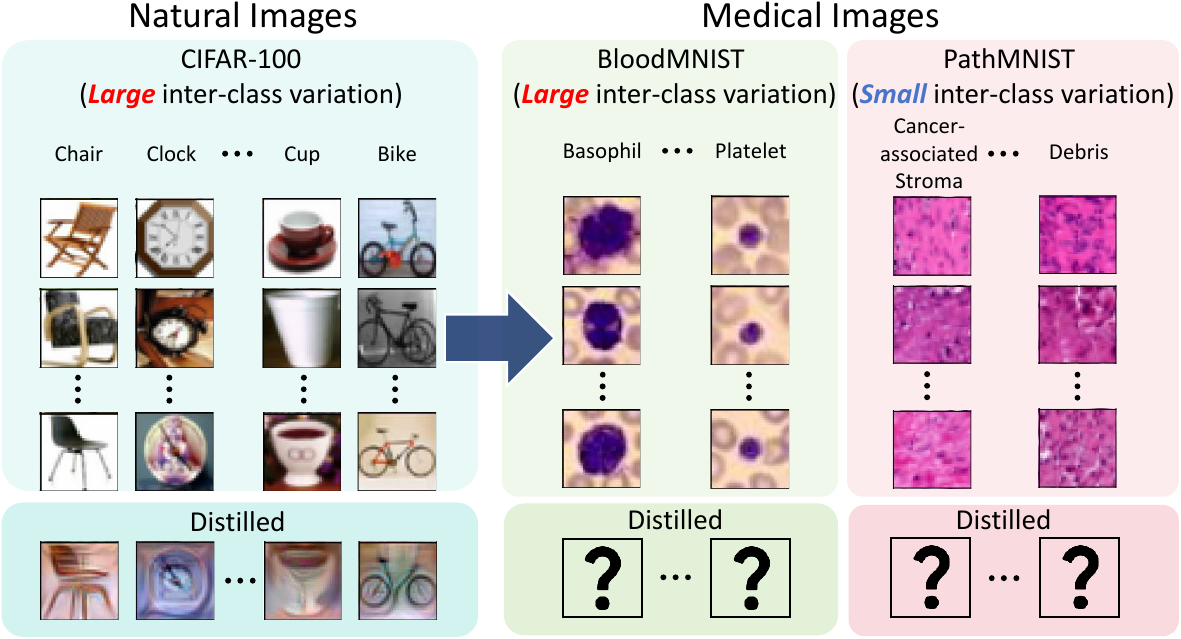}
    \caption{Difference between natural and medical datasets}
    \label{Natural}
\end{figure}

Therefore, this work focuses on exploring the data distillation of medical images in a wide range of data modalities, including colon pathology, microscope, dermatoscope images, retinal OCT and abdominal CT, for multi-class classification tasks. Initially, we aimed to address these questions as a starting point for our research:
1) How does the distillation process for medical datasets compare to established methods used for natural images in specific contexts or applications? 2) What is the performance across the 9 different medical databases (one integrated dataset and eight individual datasets)? 3)Is there any indicator to predict the distillation performance to some degree?

In this paper, we propose a universal medical dataset distillation pipeline for effective data sharing in healthcare, and we provide comprehensive experimental analysis methods to address the above three key questions. This paper will contribute to the following parts:

$\bullet$ Assess the impact of data distillation across multiple datasets characterized by minor or great variations.

$\bullet$ Explore the indicator to predict the distillation performance. 

\section{Methods}

\begin{figure}[!h]
    \centering
    \includegraphics[width=0.8\textwidth]{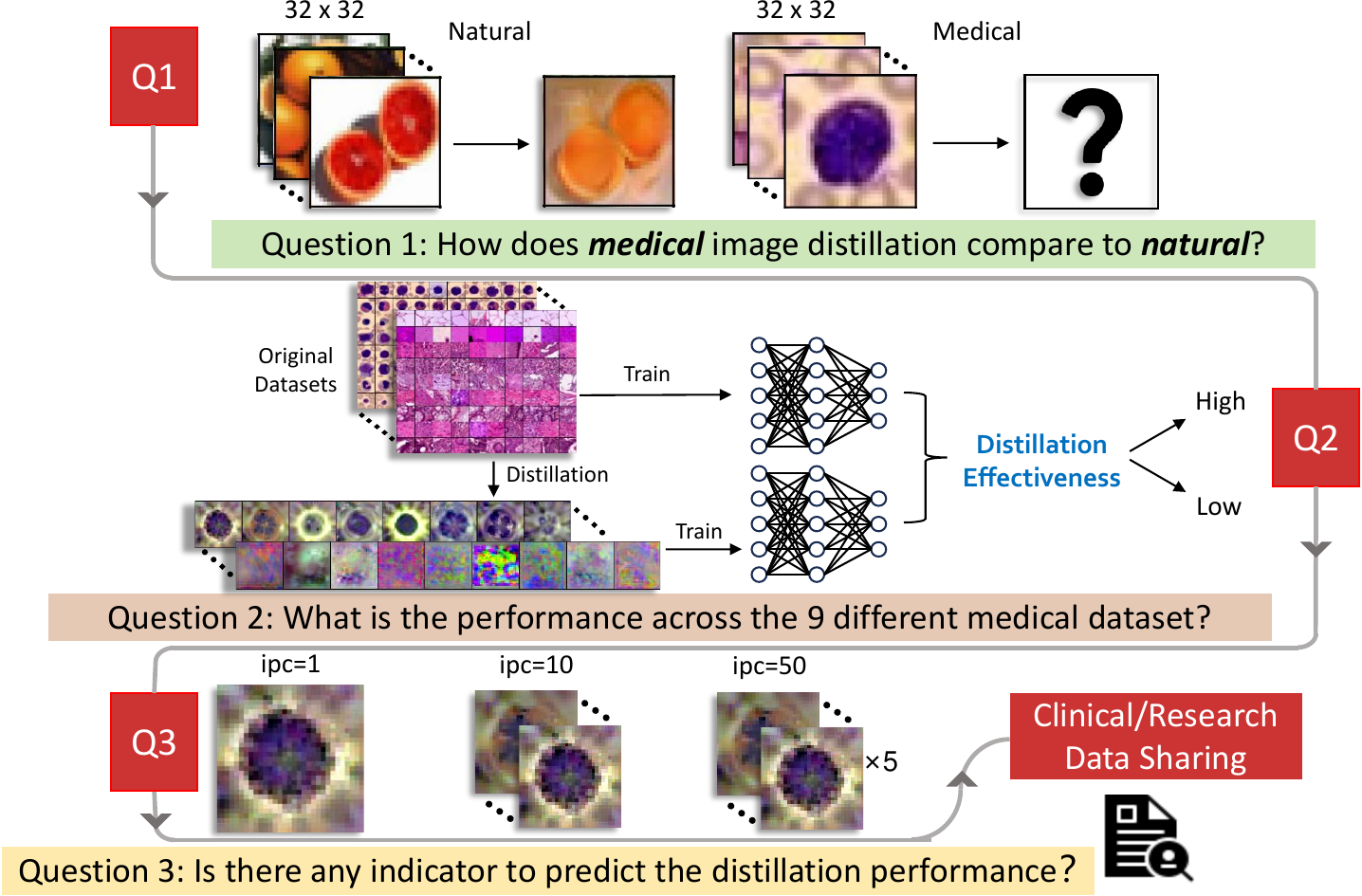}
    \caption{General working pipeline for this paper. We investigate the effectiveness of distillation on medical datasets by designing multiple experiments to answer these questions.}
    \label{pipeline}
\end{figure}

The universal medical distillation-based method is presented in Fig.~\ref{pipeline}, where we try to answer 3 questions to evaluate the performance. Specifically, we evaluate two benchmark distillation methods for classifying multimodal and multiple unimodal medical datasets. Also, we compare the performance of data distillation with random selection baselines to propose an indicator to predict the distillation performance. 

\subsection{Distillation simulation on medical datasets with different data modality}
Firstly, we extract 10$\%$ of images from each medical dataset and relabeled them with their respective dataset names rather than inter-class names to build a new dataset. The diverse data modalities of these medical datasets ensure a similarly large inter-class variation of natural images, which allows us to test the feasibility of distillation on medical datasets after modification. Secondly, we also design multiply comparison distillation experiments on different medical datasets, which allows us to monitor their possibility in distillation.
\subsection{Dataset distillation on individual medical datasets}
For distillation on individual dataset distillation, we also selected two main dataset distillation methods, DC and MTT to evaluate the distillation performance of different medical datasets.

\subsubsection{Dataset condensation (DC)}

Dataset condensation aims to create a small, synthetic dataset \(S = \{(s_i, y_i)\}\) that maintains the generalization performance of models trained on it, closely mirroring those trained on the original, larger dataset \(T = \{(x_i, y_i)\}\). The process is guided by minimizing an empirical loss function, with the primary objective formulated as 
\begin{equation}
    \theta_T = \arg\min_\theta \frac{1}{|T|} \sum_{(x,y)\in T} l(\phi_\theta(x), y)
\end{equation}

where \(l\) denotes a task-specific loss, and \(\phi_\theta\) represents the model parameterized by \(\theta\). The goal is to ensure that the model \(\phi_{\theta_S}\) trained on the condensed set \(S\) approximates the performance of \(\phi_{\theta_T}\) trained on \(T\), facilitated by the equation
\begin{equation}
    \min_S \mathbb{E}_{\theta_0 \sim P_{\theta_0}}[D(\theta_S(\theta_0), \theta_T(\theta_0))]
\end{equation} subject to 
\begin{equation}
   \theta_S(S) = \arg\min_\theta L_S(\theta(\theta_0))
\end{equation}
This approach remarkably reduces the computational resources required for training without compromising the model's generalization capability.

\subsubsection{Distillation by matching training trajectories (MTT)}
The methodology strives to align the parameters of a student network, trained on a distilled dataset, with those of teacher networks, which were trained on the original medical dataset $D$. Initially, each teacher network $T$ is trained on $D$, resulting in parameters $\{\theta_i\}^I_0$, referred to as teacher parameters. Correspondingly, parameters from training on the distilled dataset $D_c$ at each iteration $i$ are termed student parameters $\tilde{\theta}_i$. The process aims to distill original images that encourage the student network's parameters to resemble those obtained from the actual medical dataset, starting from identical initial values. In the distillation phase, student parameters $\tilde{\theta}_i$ are initialized with $\theta_i$ which is randomly selected from teacher parameters at step $i$. We then apply gradient descent updates to $\tilde{\theta}$ against the cross-entropy loss $\ell$ of $D_c$, as per the following equation:

\begin{equation}
\tilde{\theta}_{i+j+1} = \tilde{\theta}_{i+j} - \alpha\nabla\ell (A(D_c); \tilde{\theta}_{i+j}),
\end{equation}

where $j$ and $\alpha$ represent the number of gradient descent updates and the modifiable learning rate, respectively. $A$ signifies a differentiable data augmentation module that enhances distillation efficacy~\cite{zhao2021dataset2}. After the distillation process, we compare the updated student parameters $\tilde{\theta}_{i+J}$ with teacher parameters $\theta_{i+K}$, obtained after $K$ gradient updates, to calculate the final loss $L$, expressed as the normalized $L_2$ loss between these parameter sets:

\begin{equation}
L = \frac{\|\tilde{\theta}_{i+J} - \theta_{i+K}\|^2_2}{\|\theta_{i} - \theta_{i+K}\|^2_2},
\end{equation}

Minimizing $L$ and performing backpropagation through all $J$ updates refines the student network to produce an optimized distilled dataset $D_c^*$.

\section{Data and Experimental Setting}

In our investigation, we employed the dataset distillation approach across a diverse array of medical images, leveraging eight meticulously preprocessed datasets from the MedMNIST collection~\cite{Yang_2023}. These datasets encapsulate a wide range of medical imaging types, including colon pathology (PathMNIST), dermatoscopy images (DermaMNIST), retinal OCT scans (OCTMNIST), blood cell microscopy (BloodMNIST), kidney cortex microscopy (TissueMNIST), and various abdominal CT scans (OrganAMNIST, OrganCMNIST, OrganSMNIST). Each dataset was selected for its unique data modality, encompassing a total of six different modalities, and varying significantly in scale, ranging from as few as about 10,000 images to over 165,000 images for training, which resembles the diversity of dataset sizes which resembles the diversity of datasets within medical imaging research.
To accommodate our analysis framework—a simple ConvNet architecture
designed by Gidaris and Komodak~\cite{gidaris2018dynamic}—all images were resized to a uniform dimension of 32x32 pixels, slightly adjusted from their original 28x28 format. This standardization was critical for maintaining the integrity of the images while ensuring compatibility with our computational model.
To further refine our experimental setup and enhance the reliability of our findings, we applied ZCA whitening across all datasets. This preprocessing step was critical for normalizing the images, which reduced redundancy and emphasized critical features, and effectively improved the model's ability to generalize across diverse medical imaging scenarios. Through these methodological enhancements, our study sought to advance the understanding and application of dataset distillation techniques within the nuanced field of medical imaging, offering insights into the complexities and requirements specific to this domain.
For experiment settings, two distillation methods, DC and MTT, were tested on the 8 datasets from MedMNIST when IPC (images per class)$=$1, 10, and 50. The batch size for training and for real data evaluation was 256 for IPC$=$1, and 128 for IPC$=$10 or more, with a training rate of $5\times 10^{-6}$. All the experiments were run on a 16GB NVIDIA RTX5000 GPU.

\section{Results}

\begin{figure}[!h]
    \centering
    \includegraphics[width=0.8\linewidth]{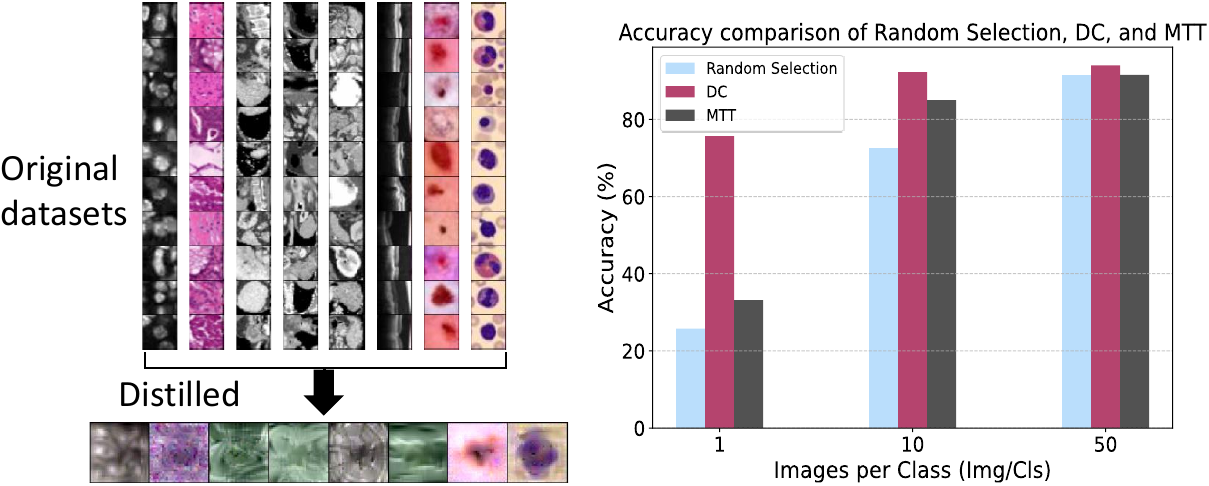}
    \caption{Distillation simulation on a large inter-class variance integrated medical dataset. The original datasets, from left to right, are TissueMNIST, PathMNIST, OrganSMNIST, OrganCMNIST, OrganAMNIST, OCTMNIST, DermaMNIST, and BloodMNIST, each treated as a single class in this integrated dataset. The right figure shows that higher distillation accuracy highlights the potential for increasing inter-class variance in specific medical datasets for improved distillation.}
    \label{MedMNIST}
\end{figure}

Fig~\ref{MedMNIST} demonstrates that the DC method consistently achieves higher accuracy rates than MTT and random selection in medical dataset distillation, suggesting its superior capability to handle the large inter-class variations characteristic of medical datasets.
Table 1 depicts a set of results comparing different methods of dataset distillation on medical images, illustrating the effectiveness of the distillation process. It shows several subsets of the MedMNIST dataset, with varying medical imaging modalities such as blood samples, skin conditions, and different organ scans. The right side of the image presents a table with accuracy percentages across random selection and two distillation methods: DC and MTT, at different numbers of images per class (1, 10, 50).
From the accuracy metrics, it is evident that both DC and MTT methods greatly outperform the random selection baseline. This implies a successful distillation that encapsulates large inter-class variation into a smaller, more manageable dataset size. For instance, with only 10 images per class, DC achieves over 90$\%$ accuracy in some cases, which suggests that the method is particularly efficient in retaining critical features from a vast, variable dataset. These results highlight the potential of dataset distillation techniques to simulate large inter-class variations effectively, allowing for the creation of condensed datasets that still carry the essential information needed for high-performance medical image analysis.

\begin{figure}[!h]
    \centering
    \includegraphics[width=0.7\linewidth]{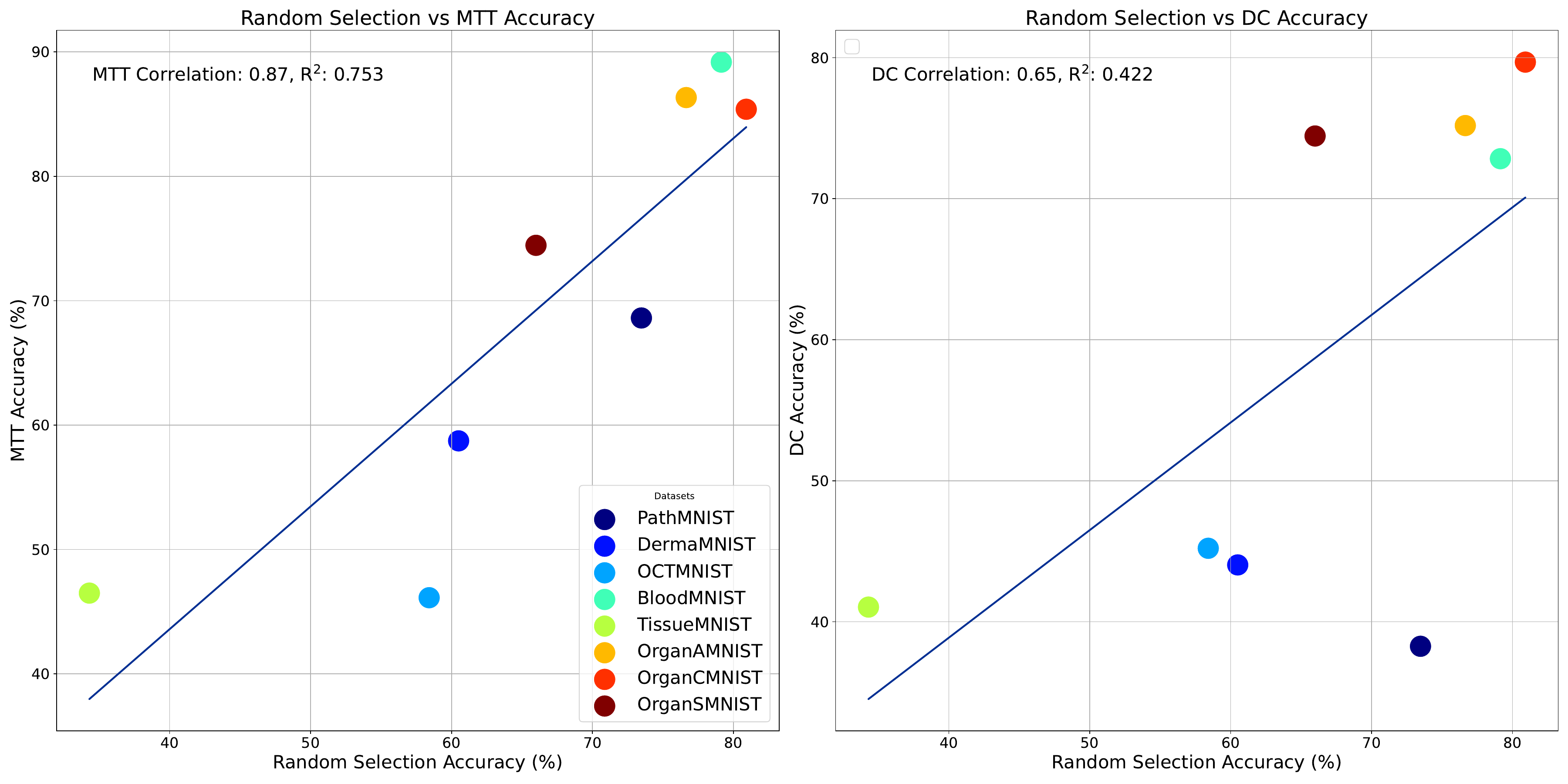}
    \caption{This figure shows the correlation between MTT/DC and Random Selected method. Comparing MTT and DC's performance on IPC$=$50 referred to randomly selected 50 images from original datasets, MTT shows higher correlation than DC, which indicates its possible better distillation performance on higher IPC. This can answer the third question in our main pipeline.}
    \label{MTT}
\end{figure}

All the accuracies on the randomly selected images from the full dataset and on the distilled images with different distillation methods are listed in Table 1. The synthetic datasets obtained from BloodMNIST, TissueMNIST, OrganAMNIST, OrganCMNIST and OrganSMNIST show performance comparable to the state-of-the-art accuracy obtained from the corresponding full datasets as expected. For PathMNIST, DermaMNIST and OCTMNIST, however, random selection method shows better performance than the other two distillation methods when IPC is 50. This indicates that the distillation method may not work well on these datasets. Fig~\ref{MTT} shows a strong linear correlation between the accuracy of the distilled dataset and the randomly selected dataset. For most of the datasets, the best accuracy is realized on 50 IPC. These experiments allow us to decide the data sharing strategy: we can first distill 50 images and then compare with 50 randomly selected medical images from full datasets. If the performance of the small synthetic dataset is better than the randomly selected, then it is likely that further distillation with a different IPC may yield both more meaningful accuracy insurance and smaller distilled dataset size.

In addition to these findings, our correlation study (Figure~\ref{MTT}) revealed a strong linear relationship between the accuracies of distilled datasets and randomly selected subsets. This suggests that the performance on a small, randomly selected sample can serve as a heuristic for predicting the effectiveness of dataset distillation. The variance in the correlation across different runs was minimal, supporting the robustness of this heuristic. This relationship likely exists because random selection acts as a proxy for task simplicity: simpler tasks with distinct feature spaces are more likely to benefit from data distillation.
\begin{figure}[!h]
    \centering
    \includegraphics[width=0.7\linewidth]{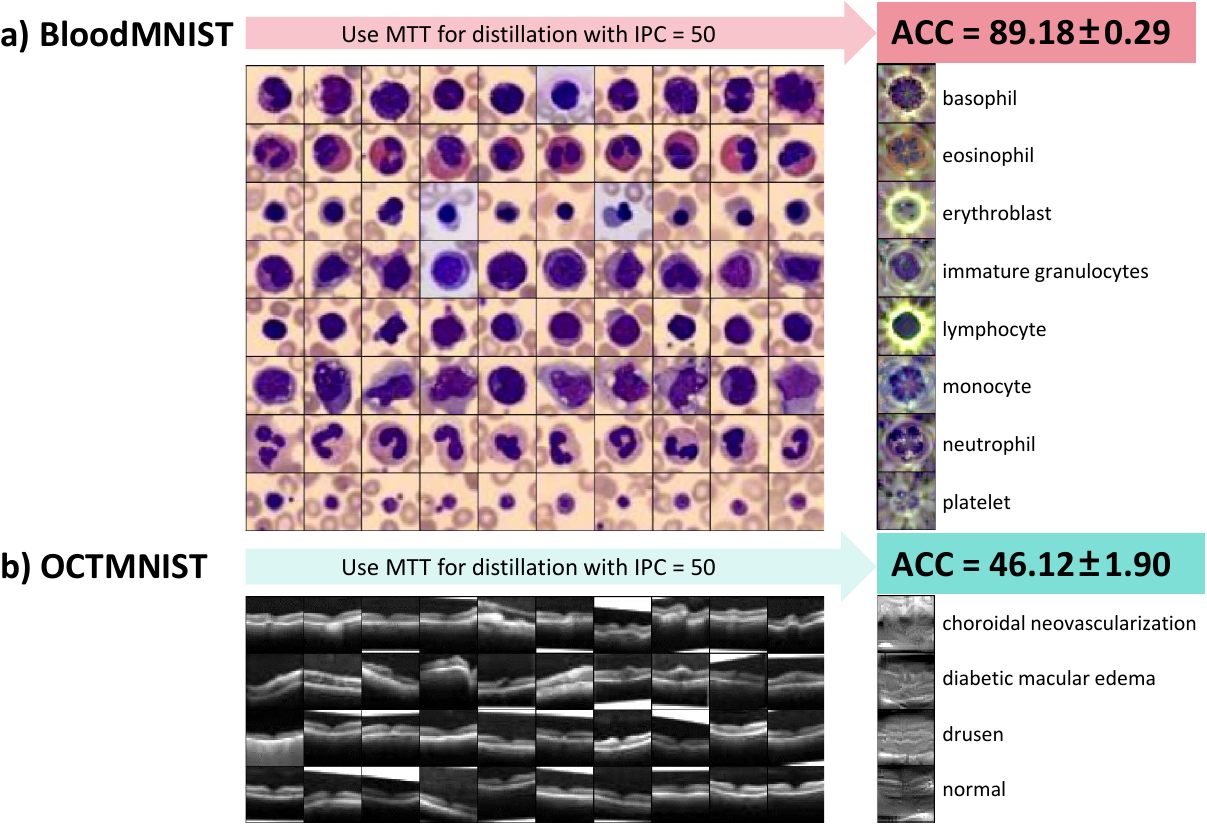}
    \caption{We test distillation methods on 9 different medical datasets, and most of large variation medical dataset shows relatively ideal distillation performance as is shown like a), and most of small variation datasets, such as b), shows lower effectiveness in distillation.}
    \label{Examples}
\end{figure}

\begin{table}[!h]
\centering
\label{table:ACC}
\scriptsize
\setlength{\tabcolsep}{2pt}
\resizebox{0.8\linewidth}{!}{%
\begin{tabular}{c|c|ccc|c}
\toprule
 Dataset & Img/Cls  & Random Selection & DC & MTT & Full Dataset \\

\midrule

\multirow{3}{*}{PathMNIST } 
& 1  & 26.02 $\pm$ 13.41 & 24.98 $\pm$ 2.74 & 13.40 $\pm$ 0.56 & \multirow{3}{*}{90.70 $\pm$ 0.10} \\
& 10   &  57.69 $\pm$ 5.68& 42.24 $\pm$ 0.89 & 32.35 $\pm$ 2.07  \\
& 50  & \textbf{73.47$\pm$4.06} & 38.26 $\pm$ 1.21 & 68.61 $\pm$ 1.24  \\

\midrule

\multirow{3}{*}{DermaMNIST }
& 1 &  18.45$\pm$ 11.97 & 28.22$\pm$2.12 & 25.52$\pm$1.75 & \multirow{3}{*}{73.50 $\pm$ 0.10} \\
& 10  & 27.97$\pm$13.71 & 44.07$\pm$1.36 &  59.67$\pm$2.00 \\
& 50  & \textbf{60.50$\pm$4.20} & 44.03$\pm$2.48 & 58.73$\pm$0.54  \\
\midrule

\multirow{3}{*}{OCTMNIST } 
& 1  & 26.09$\pm$2.91 & 29.92$\pm$0.99 & 25.40$\pm$1.57 & \multirow{3}{*}{74.30 $\pm$0.10} \\
& 10  & 39.71$\pm$4.36 & 46.78$\pm$1.05 & 35.62$\pm$2.38   \\
& 50  & \textbf{58.41$\pm$3.30} & 45.21$\pm$1.33 &  46.12$\pm$1.90 \\
\midrule

\multirow{3}{*}{BloodMNIST}
& 1  & 34.78$\pm$6.16 & 62.46$\pm$2.03 & 60.50$\pm$3.01  & \multirow{3}{*}{95.80 $\pm$ 0.10} \\
& 10  & 64.19$\pm$4.83 & 74.81$\pm$0.70 & \textbf{89.38$\pm$0.33 } \\
& 50  & 79.14$\pm$3.02 & 72.84$\pm$0.93 &  89.18$\pm$0.29 \\
\midrule

\multirow{3}{*}{TissueMNIST} 
& 1  & 23.29$\pm$6.74 & 33.83$\pm$1.91 & 13.6$\pm$1.01 & \multirow{3}{*}{67.6 $\pm$ 0.10} \\
& 10  & 28.69$\pm$2.72 & 36.25$\pm$0.77 & 35.00$\pm$1.86  \\
& 50  & 34.30$\pm$6.07 & 41.04$\pm$0.86 & \textbf{46.49$\pm$0.95}  \\
\midrule

\multirow{3}{*}{OrganAMNIST} 
& 1  & 19.05$\pm$9.44 & 48.44$\pm$0.61 & 44.04$\pm$0.53 & \multirow{3}{*}{93.50 $\pm$ 0.10} \\
& 10  & 53.03$\pm$4.30 & 75.73$\pm$0.30 & 84.52$\pm$0.47  \\
& 50  & 76.65$\pm$1.90 & 75.19$\pm$0.38 & \textbf{86.33$\pm$0.47}  \\
\midrule
\multirow{3}{*}{OrganCMNIST} 
& 1  & 25.38$\pm$6.79 & 50.04$\pm$1.54 & 67.29$\pm$1.10 & \multirow{3}{*}{90.00 $\pm$ 0.10} \\
& 10  & 57.08$\pm$4.96 & 79.03$\pm$0.22 & 84.51$\pm$0.44  \\
& 50  & 80.91$\pm$0.99 & 79.69$\pm$0.50 & \textbf{85.39$\pm$0.10 } \\
\midrule
\multirow{3}{*}{OrganSMNIST} 
& 1  & 19.67$\pm$4.67 & 32.89$\pm$1.83 & 31.17$\pm$0.68 & \multirow{3}{*}{78.20 $\pm$ 0.10 } \\
& 10   & 40.56$\pm$3.18 & 59.80$\pm$0.26 & 66.87$\pm$0.52  \\
& 50  & 65.99$\pm$1.16 & \textbf{74.45$\pm$0.46} & 69.17$\pm$0.42   \\
\bottomrule
\end{tabular}%
}
\vspace{7pt}
\caption{Comparing distillation and random selection methods in 8 different medical datasets sourced from MedMNIST. As in previous work, we distill the given number of images per class using the training set, train a neural network on the synthetic set, and evaluate on the test set. To get $\Bar{x}\pm s$, we train 5 networks from scratch on the distilled dataset. Note that the random selected method and state-of-the-art use ResNet-18~\cite{he2015deep} for all the datasets. All others use a 128-width ConvNet. The best performances of distillation or random selection method on each dataset are highlighted on bold. This table exhibits different distillation performances on medical datasets with diverse inter-class variations.}

\end{table}

\section{Conclusion}

Our study presents a comprehensive evaluation of dataset distillation techniques within the realm of medical imaging, showcasing their potential to streamline data sharing and enhance model training efficiency across diverse medical datasets. We've demonstrated that dataset distillation can effectively condense critical diagnostic information into greatly smaller datasets without compromising accuracy. Notably, our findings reveal that distillation methods excel in specific datasets, while challenges remain in others, emphasizing the need for tailored approaches. This work lays a foundational step towards optimizing data sharing strategies in healthcare, fostering advancements in medical research and patient care through more efficient and secure data utilization. Our research not only advances the understanding of dataset distillation's applicability but also opens avenues for future exploration in optimizing distillation processes for medical imaging data.

\acknowledgements
\begin{flushleft}
This research was supported by NIH R01DK135597(Huo), DoD HT9425-23-1-0003(HCY), NIH NIDDK DK56942(ABF). This work was also supported by Vanderbilt Seed Success Grant, Vanderbilt Discovery Grant, and VISE Seed Grant. This project was supported by The Leona M. and Harry B. Helmsley Charitable Trust grant G-1903-03793 and G-2103-05128. This research was also supported by NIH grants R01EB033385, R01DK132338, REB017230, R01MH125931, and NSF 2040462. We extend gratitude to NVIDIA for their support by means of the NVIDIA hardware grant. This works was also supported by NSF NAIRR Pilot Award NAIRR240055.
\end{flushleft}
\bibliographystyle{spiebib} 
\bibliography{report.bib}

\end{document}